# Monte Carlo simulations study of the intermetallic compound NdCo₂Si₂ Magnetic properties

**R. KHALLADI[1], S. IDRISSI[1], S. MTOUGUI[1], I. EL HOUSNI[1], S. ZITI[2], H. LABRIM[3], N. EL MEKKAOUI[1] , L. BAHMAD[1,*]**

[1] Laboratoire de la Matière Condensée et des Sciences Interdisciplinaires (LaMCScI), Mohammed V University of Rabat, Faculty of Sciences, B.P. 1014 Rabat, Morocco.

[2] Intelligent Processing and Security of Systems, Mohammed V University of Rabat, Faculty of Sciences, B.P. 1014 Rabat, Morocco.

[3] USM/DERS/Centre National de l'Energie, des Sciences et des Techniques Nucléaires (CNESTEN), Rabat, Morocco.

## Abstract

Magnetic properties of the intermetallic compound NdCo2Si2 are investigated by using the Monte Carlo simulation (MCs) under Metropolis algorithm. The magnetism of the compound is caused by the existence of the rare earth ($Nd^{3+}$) ions with a magnetic moment taking the value 2.7 µB. Firstly, the ground state phase diagrams are presented and discussed in different planes corresponding to different physical parameters of the system. The stable phases are explored for different configurations of the Hamiltonian of the system. These stable phases are determined by the minimal energies. For non-null temperature values, we compute the magnetizations and susceptibilities behaviors as a function of temperature by using the Monte Carlo simulations (MCS). Also, we present the magnetization behaviors as a function of the exchange coupling interactions, the crystal field and the external magnetic field. Finally, we present and discuss the magnetic hysteresis loops of the intermetallic NdCo₂Si₂ compound as a function of the external magnetic field for fixed values of temperature and the other physical parameters.

## Key words:

Magnetic properties; Intermetallic compound; Rare earth; hysteresis loops; NdCo₂Si₂; Monte Carlo simulations.

*) Corresponding authors: khalladirajaa17@gmail.com (K.R.) ; lahou2002@gmail.com (L.B.).

# I.    Introduction

Recently, many experimental and theoretical investigations have been done in the field of the development of materials and solid state science. This progress is reflected in different domains such as economic, social and scientific etc. Among the topics most studied in solid states we found nano-materials, spintronic materials, graphene and intermetallic compounds [1, 2]. Generally this family of rare earth based intermetallic materials includes minimum two metallic elements in their composition, their crystal structures are ordered and shows a rich variety of interesting physical properties [3-6]. The basic form of this series is RMX where the R represents the rare earth and the M stands for the transition metals elements in the periodic table and the X is a non-magnetic element. Many researchers have extracted several properties of this type of materials ranging from superconductivity [7], nearly ferromagnetic Fermi-liquid, heavy fermion behavior and complex magnetic ordering states [8, 9], magnetoresistance (MR) [10]. This latter property is very studied because of its utilities in sensors and magnetic memory devices also in high density read head technology [11] which make it an interesting subject of various investigations [12-16].

The $NdCo_2Si_2$ compound which is part of ternary rare earth based intermetallic family crystallizes in $ThCr_2Si_2$-type structure where the plans are stacked in the sequence Nd-Si-Co-Si-Nd along the c axis [17]. Also, it has been found that this compound crystallizes anti-ferromagnetically bellow $T_N \approx 32$ K [18].On the other hand, it has been noted a ferromagnetic alignment of the moments in the layers perpendicular to the c-axis. The magnetism in this compound is due to the Nd atoms ($Nd^{3+}$).

This family of intermetallic compound series of $RT_2X_2$ was the goal of several investigations in many interests such as developing high-temperature in superconductors [19-23]. Physical properties are also extracted of the ternary intermetallic $CaFe_2As_2$ compound like electronic and magnetic properties [23, 24]. John T. Sypek *et al* have been reported the superelastic behavior in this intermetallic compound [25]. In addition, the series of the $RECo_2Si_2$ (RE = rare earth) intermetallic compounds are extensively studied where they are focused on the magnetoresistance (MR) phenomena [26-30].

The aim of this work is to investigate the magnetic properties of the intermetallic compound $NdCo_2Si_2$ using Monte Carlo simulations under Metropolis algorithm. Also we provide the magnetic behavior of the total magnetizations as a function of temperature and physical parameters. This paper is organized as follows: In part II, we provide the model and the used theory. In section III, we present and discuss the obtained results. Section IV, is devoted to the conclusions.

## II. Model and theory

The trigonal structure of the intermetallic NdCo2Si2 (see fig.1.a) is constituted of three type of atoms: 'Nd' which occupies a unique crystallographic site (1a), while both Cobalt 'Co' and Silicium 'Si' are placed in the two sites (4k), (0) for 'Co' and (8r), (0) for 'Si', respectively (see table.1) [31, 32].

Since the studied compound (NdCo2Si2) is magnetic, the only magnetic atom with important moment is the rare earth element Neodym (Nd).This compound is widely identified by the Neodym ($Nd^{3+}$) ions having the spin moment S=5/2, with an effective magnetic moment $\mu_{eff}$=2.7 $\mu_B$. We propose the following Hamiltonian in order to describe the studied system:

$$\mathcal{H} = -J_1 \sum_{<i,j>} S_i\, S_j - J_2 \sum_{<k,l>} S_k\, S_l - H \sum_i S_i - D \sum_i S_i^2 \qquad (1)$$

With $S_i=\pm5/2; \pm 3/2; \pm 1/2$. Where: the notations $<...>$ stand for the summations over the first and second Nd ions with the exchange coupling interactions $J_1$ and $J_2$, respectively (see Fig.1.b). The crystal field is denoted D and H stands for the external magnetic field.

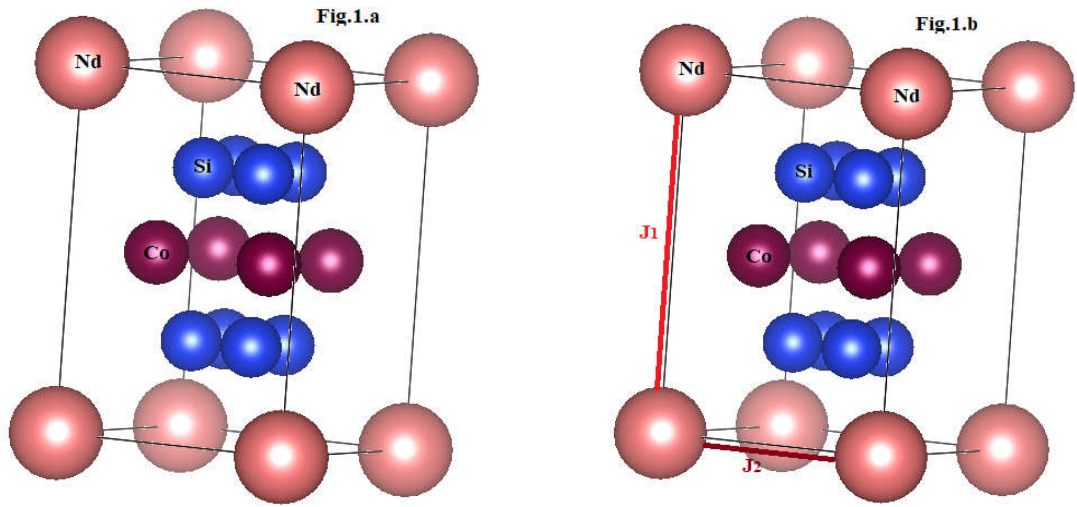

**Fig.1.** The crystalline structure of the intermetallic NdCo₂Si₂ compound. Where the blue spheres represent the silicium 'Si' atoms, the violet spheres stand for the cobalt 'Co' atoms (a). The only exchange coupling interactions Nd-Nd in the super-cell unit (b).

| atoms | X | Y | Z | Occupancy | Site |
|-------|-----|-----|-----|-----------|------|
| Nd | 0.0000 | 0.0000 | 0.0000 | 1 | 1a |
| Si | 0.36710 | 0.36710 | 0.73420 | 1 | 8r |
| Si | 0.63290 | 0.63290 | 0.26580 | 1 | 0 |
| Co | 0.25000 | 0.75000 | 0.50000 | 1 | 4k |
| Co | 0.75000 | 0.25000 | 0.50000 | 1 | 0 |

**Table1**. Coordinates of the NdCo$_2$Si$_2$ compound [32]

### III.     Monte Carlo simulation study:

In order to study the magnetic properties of the studied NdCo$_2$Si$_2$ compound, we perform Monte Carlo simulations under the Metropolis algorithm. This method is based on the Markov process using the Hamiltonian of Eq. 1. We start from initial configurations to reach the equilibrium point of the studied system. At each spin configuration, a number of $10^5$ Monte Carlo steps are performed; discarding the first $10^4$ generated ones. At each Monte Carlo step, the all sites in the system are swept and single-spin flip attempts are made. The principle of the Metropolis algorithm is to accept or reject the exchanges in the spin configurations. The periodic boundary conditions are applied for the finite lattice size. Thus, the calculation of the physical parameters such as the magnetization, the magnetic susceptibility, the specific heat, energy of the system is performed. Such parameters are calculated with the following expressions:

The total energy of the system per site is:

$$E_{tot} = \frac{1}{N} < \mathcal{H} > \quad (2)$$

Where N denotes the total number of magnetic atoms in the supercell unit.

The total magnetization of the system:

$$M_{tot} = \frac{1}{N} < \sum_i S_i > \quad (3)$$

The total susceptibility and specific heat of the system are:

$$\chi_{tot} = \frac{\beta}{N}\left(\langle M_{tot}{}^2 \rangle - \langle M_{tot} \rangle^2\right) \quad (4)$$

$$c_v = \frac{\beta^2}{N}\left(\langle E_{tot}{}^2 \rangle - \langle E_{tot} \rangle^2\right) \quad (5)$$

With: $\beta = \frac{1}{K_B T}$, where $K_B$ is the Boltzmann constant assumed to take its unit value, T is the absolute temperature.

## III.    Results and discussions

Based on the equation of the Hamiltonian (1), we compute the energies of all possible $6=2\times5/2+1$ configurations. For this purpose, we plot in Fig.2 the corresponding ground state phase diagram for D=0 and $J_2$=1 in the plane (H, $J_1$). From this figure, it is found that the all possible configurations are stable, namely: ±5/2, ±3/2 and ±1/2. A perfect symmetry is found in this figure. Also, the phases with maximum magnetic moment are found to be stable only for values of the parameter J1≥-5. The other phases are stable only for values of the parameter J1< -5. The same topology has been found in the plane (H, J2) for J1=1 and D=0 that is why we did not give this figure.

On the other hand, to inspect the crystal field effect on the stable configurations, we illustrate in Fig.3 the obtained results of the ground state phase diagram in the plane (D, J1) for H=1, D=0 and J2=1. Surprisingly, only the two phases -5/2 and -1/2 are stable in this plane. This is probably due to competition between the applied external magnetic field and the exchange coupling interaction, also to the antiferromagnetic behavior of this system. Also, we found an analogue topology with this figure when we study the effect of J2 in the plane (D, J2) with the same stable phases.

For non-null temperature values, we perform Monte Carlo simulations in order to simulate the magnetic properties of the intermetallic compound NdCo2Si2. In fact, we provide in Fig.4.a the obtained total magnetizations as a function of temperature for $H=J_1=J_2$=1 and several values of the crystal field: D=-2, 0, 3. From this figure, it is found that the decreasing crystal field effect is to reach rapidly the saturation of the total magnetization. This can be explained by the antiferromagnetic behavior of this system. The corresponding susceptibilities as a function of temperature are presented in Fig.4.b for the same values as Fig.4.a: $H=J_1=J_2$=1 and D=-2, 0, 3. It is found that the peak of each susceptibility curve is displaced towards higher temperature when the value of the crystal is increased. This is in good agreement with the behavior of the total magnetizations shown in Fig.4.a.

In order to underline the effect of the crystal field on the total magnetizations, we give in Fig.5 the obtained results as a function of the crystal field, in (a) for T=20 K, J1=J2=1 and selected values of external magnetic field H=0, 1 and -1; in (b) for H= J1=J2=1 with variation of temperature T=15, 20 and 30 K and in (c) for T=20 K, H=1, and several values of the exchange coupling interactions (J1=1, J2=1), (J1=-1, J2=1), (J1=-1, J2=-1) and (J1=1, J2=-1).

It is found that for D< -4, there is no external magnetic field effect on the total magnetizations as it is illustrated in Fig. 5.a. For D>-4, the saturation sign of the total magnetizations follows the sign of the external magnetic field. In fact, for H=1 the total magnetizations reach their saturation with positive value. While for H=0 or -1, the total magnetizations reach their saturation with negative value. From Fig.5.b, it is seen that for D<-4 there is no temperature effect on the total magnetizations; while for D>-4 the saturation of the total magnetization is reached with a first order transition. In addition, the increasing temperature effect is to decrease the saturation value of the total magnetizations, see Fig.5.b for T=15, 20 and 30K.

The effect of varying the exchange coupling interactions is illustrated in Fig.5.c. Indeed, from this figure, it is found that for (J1=1, J2=1) a maximum saturation value of the total magnetization is reached when increasing the crystal field values. For (J1=1, J2=-1) a saturation value of about +0.5 is reached when the crystal field increases; while for (J1=-1, J2=1) and (J1=-1, J2=-1) the saturation of the magnetization values is reached rapidly when increasing the crystal field parameter values.

With the intention of showing the effect of varying the exchange coupling interaction J1 on the behavior of the total magnetizations, we present in Figs.6 (a, b and c) the obtained results. Effectively, we provide in Fig.6.a the variation of the total magnetizations as a function of the exchange coupling $J_1$ for $J_2$=1, H=D=0, T=15 K, T=20 K and T=35 K. It is found that for J1< -1.5, there is no temperature effect on the total magnetizations as it is illustrated in Fig 6.a. For J1>-1.5, the total magnetizations reached its saturation with positive value for T=35 K; while this saturation is reached with negative values for T=15 and 20 K. The effect of varying the external magnetic field is presented in Fig.6.b for T=20 K, D=0and selected values of the external magnetic field: H=-1, 0 and 1. From this figure, we can distinguish three regions: (i) for J1< -7.5 the total magnetizations are not affected by the variation of the external magnetic field; (ii) in the region -7.5 <J1< 0 the total magnetization undergoes two transitions of first order type before reaching its saturation. In the region (iii) where J1>0, the total magnetizations do not depend on the increasing of the external magnetic field and reach a negative value of their saturation. To show the effect of varying the crystal field on the behavior of the total magnetizations we present in Fig.6.c, the obtained results for T=20 K, H=0 and selected values of the crystal field D=1, 0 and -1. One can distinguish two regions: (i) for J1<0 the total magnetizations are practically independent on the variation of the crystal field. For the region (ii) where J1>0, undergo a first-order transition before reaching their saturation value. This saturation value follows the sign of the fixed crystal field value.

To see the effect of varying the parameter J2 on the behavior of the total magnetizations, we plot in Figs.7 (a, b and c) the obtained results when fixing the other physical parameters. In fact, Fig.7.a summarizes the obtained curves as a function of J2, for $J_1=1$, H=1, D=0 and fixed temperature values T=10, 20 and 40 K. From this figure, one can depict that for J2<0 the total magnetizations are weakly affected by the increasing temperature effect. For J2>0, the total magnetizations undergo a first order transition before reaching the saturation value. The competition between the increasing temperature effect and the exchange coupling interaction as well as the antiferromagnetic behavior are present in this figure. The same advanced arguments in Fig.7.a are still valid in Fig.7.b, plotted for T=20 K, D=0, when varying the external magnetic field when taking the values: H=-1, 0 and 1. To inspect the effect of varying the crystal field when increasing the exchange coupling interaction J2, we present in Fig.7.c, the obtained results for T=20 K, H=1 and selected values of crystal field D=-1, 0 and 1. The same behavior of the total magnetizations is found in this figure when varying the crystal field, except for the negative values of this parameter. When we increase the parameter J2, the all magnetizations reach their saturation after a first order transition.

To complete this study, we give in Figs. 8a and 8b, the hysteresis loops of the total magnetizations for the intermetallic compound NdCo2Si2 for a fixed temperature T=20 K. The variation of the crystal field on such hysteresis loops is illustrated in Fig.8.a for J1=J2=1 and the selected values of the crystal field D=-5, 0 and 7. From this figure, the increasing crystal field effect is to increase the surface of the loops as well as the corresponding coercive field in the case where the crystal field takes positive values. While for the negative values of this parameter, the hysteresis cycle shows multiple secondary loops. This is due to the competition between different physical parameters including the temperature the exchange coupling interaction and the crystal field. The effect of varying the exchange coupling interactions on the hysteresis loops is presented in Fig.8.b in the absence of the crystal field D=0. For (J1=1, J2=1), the corresponding surface loop is maximum when compared with the other values. When the exchange coupling interactions are chosen so that (J1=-1, J2=1) we are witnessing to the appearance of intermediate states. In addition for the specific case where (J1=1, J2=-2) the apparition of secondary loops is caused by the antiferromagnetic behavior of the NdCo2Si2 compound.

Finally, Fig. 9 illustrates the behavior of the coercive field as a function of the crystal field for a fixed temperature T = 20 K and the exchange coupling interaction values J1=J2= 1. From this figure it is seen that the coercive field decreases linearly when increasing the crystal field. This decreasing is caused by the competition between different physical parameters namely: the temperature, the exchange coupling interactions and the crystal field.

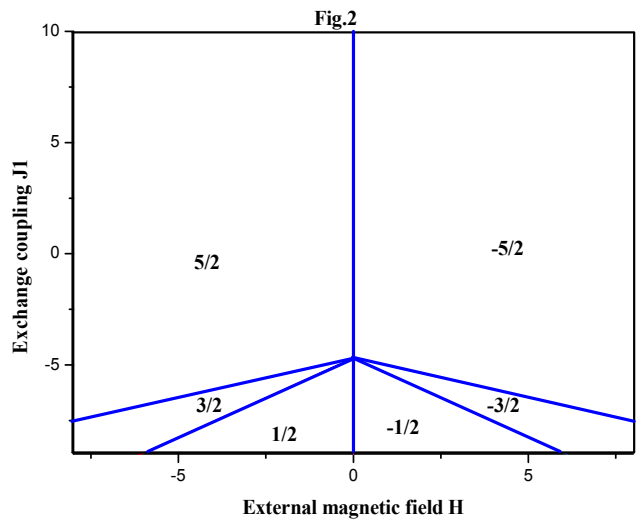

**Fig.2.** Ground state phase diagram for D=0 and $J_2$=1, in the plane (H, $J_1$).

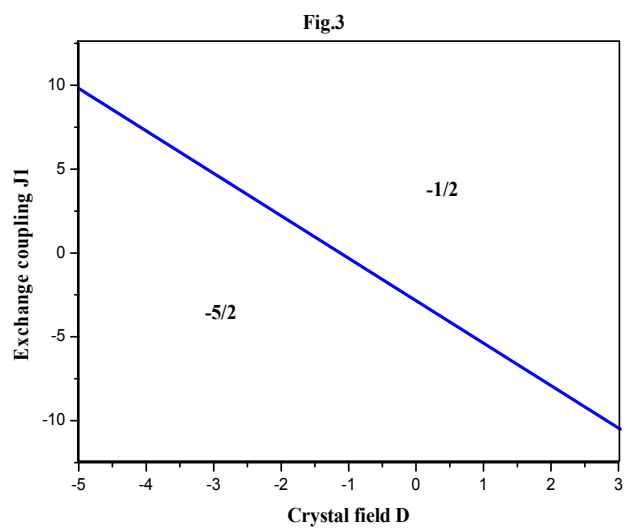

**Fig.3.** Ground state phase diagram in the plane (D, J1) for H=1, D=0 and J2=1.

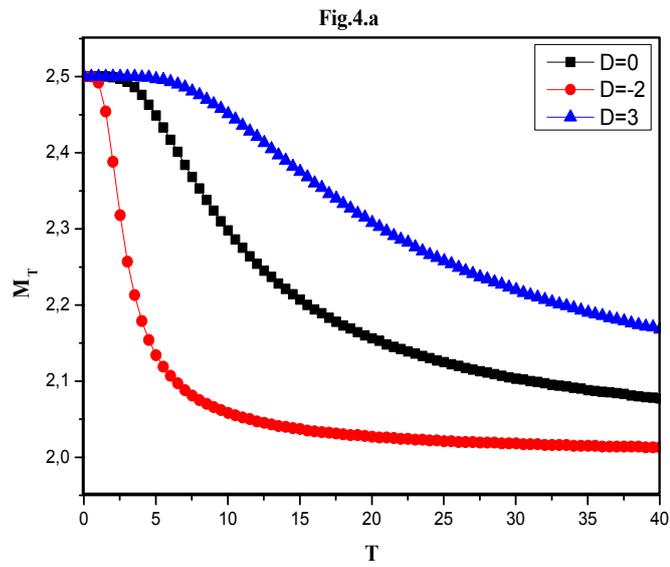

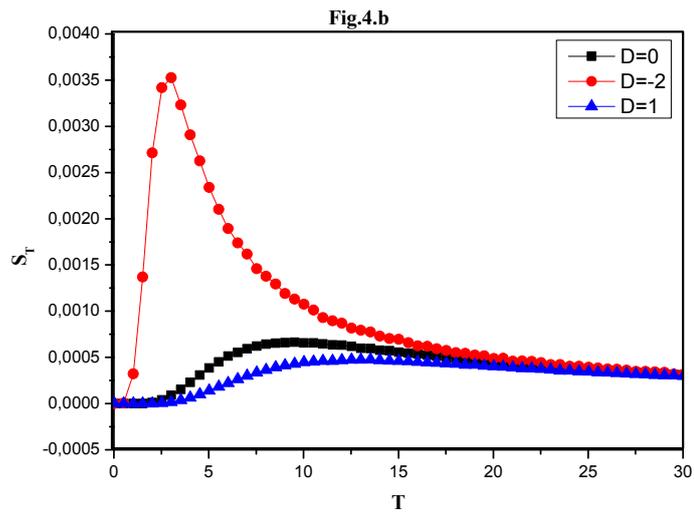

**Fig.4** (a) total magnetizations and (b) total susceptibilities as a function of temperature for H=J$_1$=J$_2$=1 and D=-2, 0, 3.

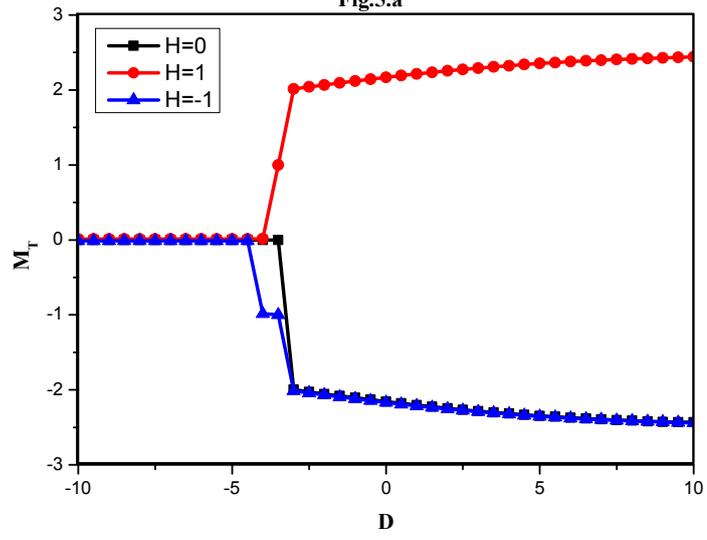

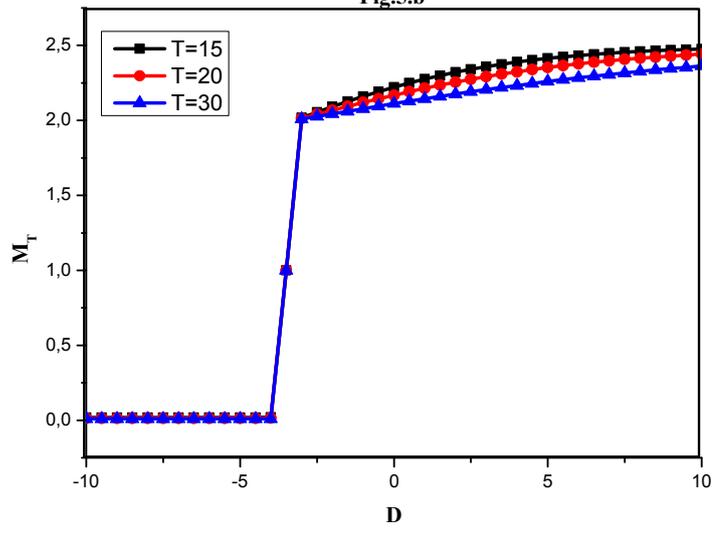

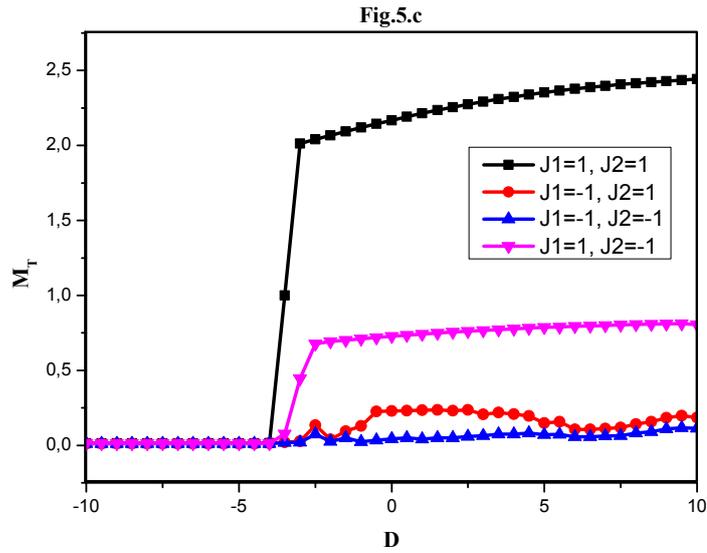

**Fig.5.** variation of the magnetization as a function of the crystal field (a) for T=20 K, J1=J2=1 and selected values of external magnetic field H=0, H=1 and H=-1; (b) for H= J1=J2=1 with variation of temperature T=15 K, T=20 K and T=30 K and (c) for T=20 K, H=1, and fixed values of the exchange coupling interactions.

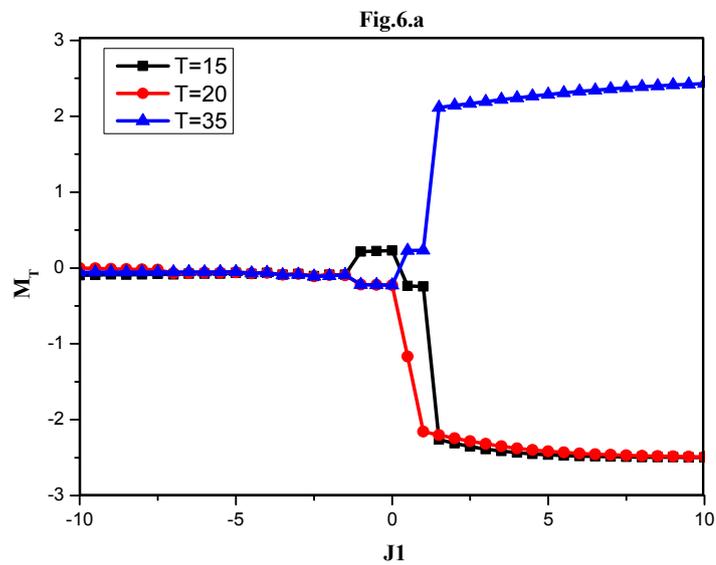

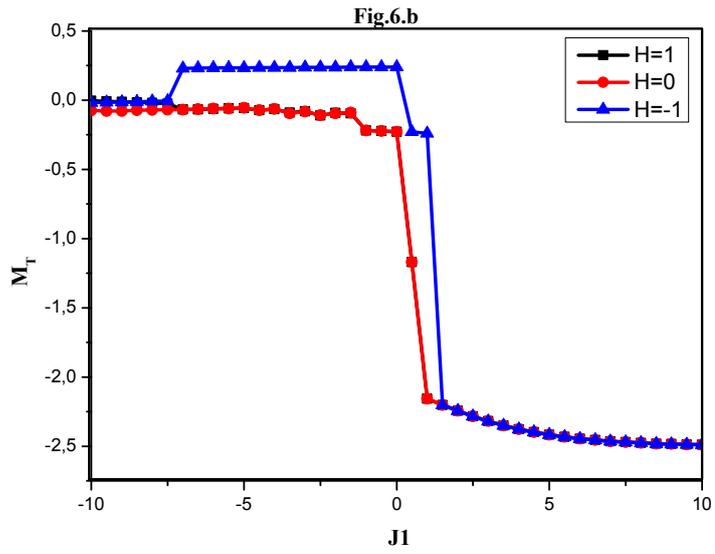

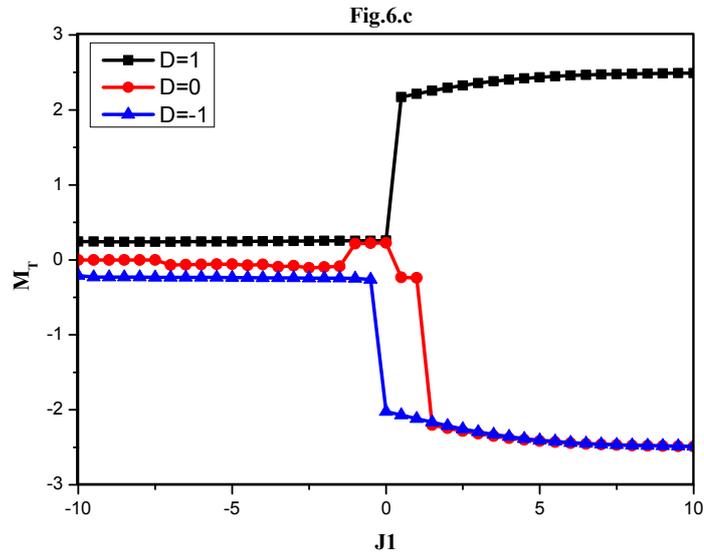

**Fig.6.** Magnetization as a function of the exchange coupling $J_1$ for $J_2=1$ (a) for H=D=0, T=15 K, T=20 K and T=35 K; (b) for T=20 K, D=0 and selected values of the external magnetic field H=-1, 0 and 1; and (c) for T=20 K, H=0 and selected values of the crystal field D=1, 0 and -1.

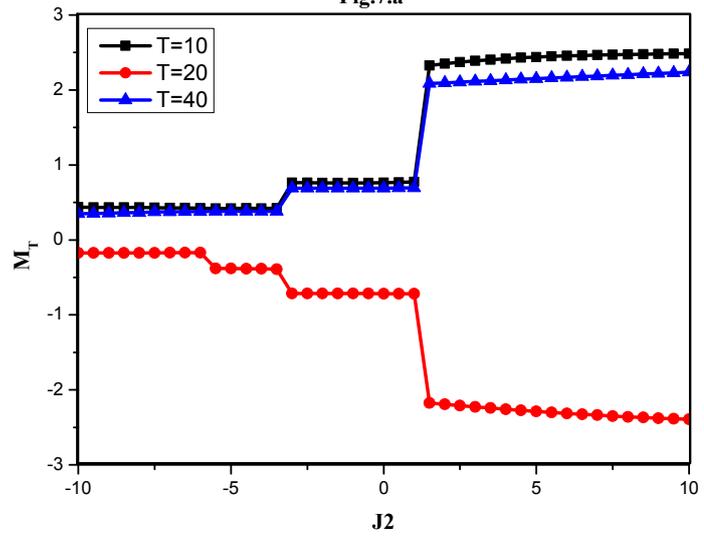

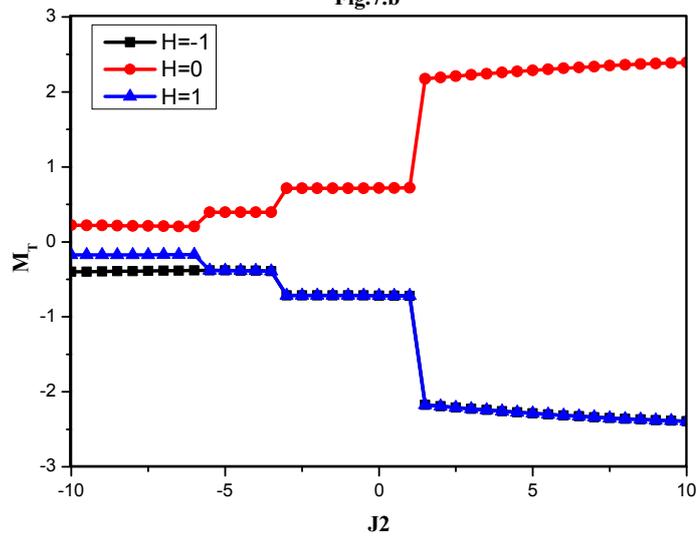

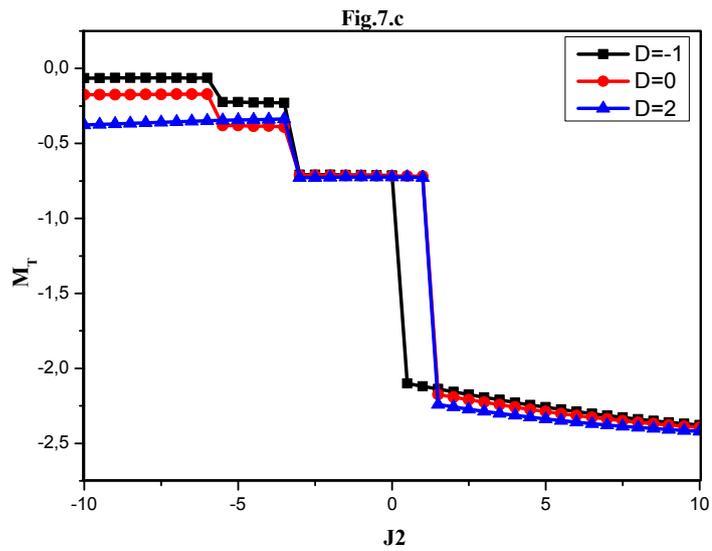

**Fig.7.** Profile of the total magnetizations as a function of the exchange coupling $J_2$ for $J_1=1$; in (a) for H=1, D=0 and fixed temperature values T=10, 20 and 40 K; in (b) for T=20 K, D=0 and selected values of the external magnetic field H=-1, 0 and 1; and in (c) for T=20 K, H=1 and selected values of crystal field D=-1, 0 and 1.

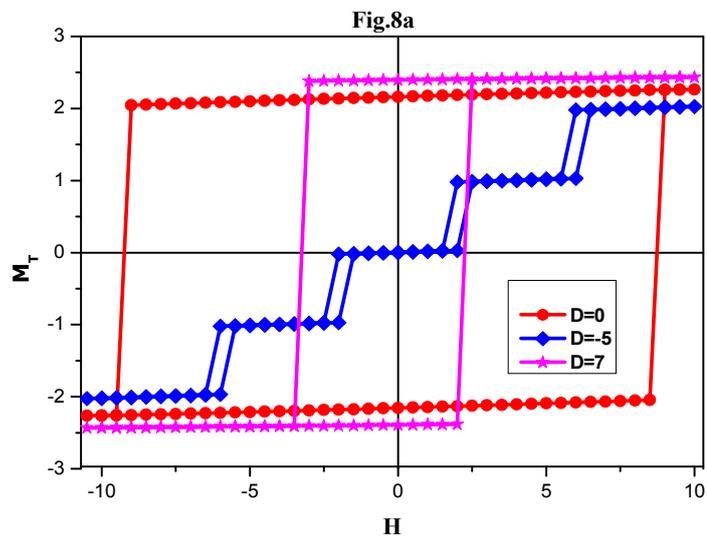

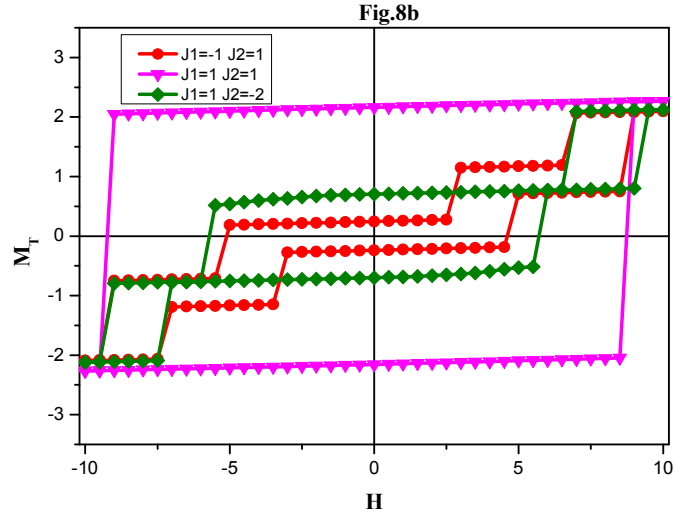

**Fig.8.** The hysteresis loops for the $NdCo_2Si_2$ compound for T=20 K, (a) for J1=J2=1 and selected values of the crystal field D=-5, 0 and 7, (b) for D=0 and different values of the exchange coupling $J_1$ and $J_2$ (J1=-1, J2=1), (J1=1, J2=1) and (J1=1, J2=-2).

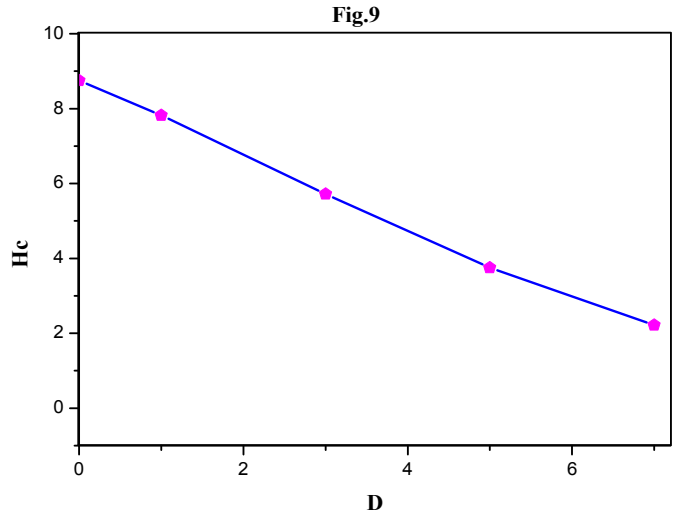

**Fig. 9.** Coercive field as a function of the crystal field for T = 20 K and J1=J2= 1.

## IV. Conclusion

In this work we have studied the magnetic properties of the intermetallic compound NdCo2Si2 by using the Monte Carlo simulation (MCs) under Metropolis algorithm. It has been shown that the magnetism of the compound is caused by the existence of the rare earth (Nd$^{3+}$) ions with a magnetic moment taking the value 2.7 µB. In a first step we have presented and discussed the ground state phase diagrams corresponding to different physical parameters of the system. It is found that the all possible configurations are stable, namely: ±5/2, ±3/2 and ±1/2 and a perfect symmetry is found in the plane (H, J1).

These stable phases are determined by the minimal of energies. For non-null temperature values, we perform Monte Carlo simulations to compute the magnetizations and susceptibilities behaviors as a function of temperature. When plotting the total magnetizations as a function of the crystal field it is found that for D< -4, there is no external magnetic field effect on the total magnetizations. While, for D>-4, the saturation sign of the total magnetizations follows the sign of the external magnetic field. In fact, for H=1 the total magnetizations reach their saturation with positive value. While for H=0 or -1, the total magnetizations reach their saturation with negative value. Finally, we present and discuss the magnetic hysteresis loops of the studied intermetallic NdCo$_2$Si$_2$ compound as a function of the external magnetic field for fixed values of temperature and other physical parameters.

# V.    References

[1] R. Khalladi, A. Benyoussef, L. Bahmad, Physica A,Volume 509, (2018), Pages 971-981.https://doi.org/10.1016/j.physa.2018.06.101

[2] R. Khalladi, S. Mtougui, S. Idrissi, L. Bahmad, S. Ziti, H. Labrim, Chinese Journal of Physics 56(6) (2018), pp. 2937-2947

[3] C.R. Barrett, A.S. Tetelman, W.D. Nix, the Principles of Engineering Materials, Prentice-Hall, New Jersey, 1973.

[4] A.M. Russell, Adv. Eng. Mater. 5 (2003) 629–639.

[5] S. Gupta, K.G. Suresh, A.K. Nigam, Y.V. Knyazev, Y.I. J. Phys. D: Appl. Phys. (2014), p. 365002CrossRef

[6]L. Li, O. Niehaus, M. Kersting, R. Appl. Phys. Lett. (2014), p. 924161–5

[7] Facio, Jorge I. and Betancourth, D. and Pedrazzini, Pablo and Correa, V. F. and Vildosola, V. and Garcia, D. J. and Cornaglia, Pablo S., Phys. Rev. B, **vol. 91** issue 1, (2015) pages 014409.

[8] S. Jia, S. L. Bud'ko, G. D. Samolyuk, and P. C. Canfield, Nat. Phys. 3, 334 (2007).

[9] S. Jia, Ni Ni, S. L. Bud'ko, and P. C. Canfield, Phys. Rev B 80, 104403 (2009).

[10] R. Roy Chowdhury, S. Dhara, I. Das, B. Bandyopadhyay, R. Rawat. Journal of Magnetism and Magnetic Materials 451 (2018) 625–628

[11] R. B. van Dover, E. M. Gyorgy, R. J. Cava, J.J. Krajewski, R. J. Felder, W. F. Peck, Phys. Rev. B47 (1993) 6134.

[12] C. Mazumdar, A.K. Nigam, R. Nagarajan, L.C. Gupta, C. Godart, B.D. Padalia, G. Chandra, R. Vijayaraghavan, Phys. Rev. B54 (1996) 6069.

[13] C. Mazumdar, A.K. Nigam, R. Nagarajan, C. Godart, L.C. Gupta, B.D. Padalia, G. Chandra, R. Vijayaraghavan, Appl. Phys. Lett. 68 (1996) 3647.

[14] E. Sampathkumaran, P.L. Paulose, R. Mallik, Phys. Rev. B54 (1996) R3710.

[15] R. Mallik, E.V. Sampathkumaran, P.L. Paulose, Appl. Phys. Lett. 71 (1997) 2385.

[16] R. Nirmala, S.K. Malik, A.V. Morozkin, Y. Yamamoto, H. Hori, Europhys. Lett. 76 (2006) 471.

[17] T. Shigeoka, N. Iwata, Y. Hashimoto, Y. Andoh, H. Fujii, J. Physique Colloq. 49 (1988) C8431.

[18] T. Shigeoka, N. Iwata, Y. Hashimoto, Y. Andoh, H. Fujii, J. Physique Colloq. 49 (1988) C8431.

[19] ] M. S. Torikachvili, S.L. Bud'ko, N. Ni, P.C. Canfield, S.T. Hannahs, Phys. Rev. B 80 (2009), 014521. .

[20] M. S. Torikachvili, S.L. Bud'ko, N. Ni, P.C. Canfield, Phys. Rev. Lett. 101 (2008), 057006.

[21] P. C. Canfield, S.L. Bud'ko, N. Ni, A. Kreyssig, A.I. Goldman, R.J. McQueeney, M.S. Torikachvili, D.N. Argyriou, G. Luke, W. Yu, Physica C 469 (2009) 404–412.

[22] N. Ni, S. Nandi, A. Kreyssig, A.I. Goldman, E. D. Mun, S.L. Bud'ko, P. C. Canfield, Phys. Rev. B 78 (2008), 014523. .

[23] S.R. Saha, N. P. Butch, T. Drye, J. Magill, S. Ziemak, K. Kirshenbaum, P. Y. Zavalij, J. W. Lynn, J. Paglione, Phys. Rev. B 85 (2012), 024525. .

[24] M. P. Allan, T.-M. Chuang, F. Massee, Y. Xie, N. Ni, S.L. Bud'ko, G.S. Boebinger, Q. Wang, D.S. Dessau, P.C. Canfield, M.S. Golden, J.C. Davis, Nat. Phys. 9 (2013) 220–224.

[25] John T. Sypek, Christopher R. Weinberger, Sriram Vijayan, Mark Aindow, Sergey L. Bud'ko, Paul C. Canfield, Seok-Woo Lee. Scripta Materialia 141 (2017) 10–14

[26] ] C. Mazumdar, A. K. Nigam, R. Nagarajan, C. Godart, L. C. Gupta, B.D. Padalia, G. Chandra, R. Vijayaraghavan, Appl. Phys. Lett. 68 (1996) 3647.

[27] L. Morellon, J. Stankiewicz, B. Garcia-Landa, P.A. Algarabel, M.R. Ibarra, Appl. Phys. Lett. 73 (1998) 3462.

[28] L. Morellon, P. A. Algarabel, C. Magen, M.R. Ibarra, J. Magn. Magn. Mater. 237 (2001) 119.

[29] S. Gupta, K.G. Suresh, A.K. Nigam, J. Alloys Comp. 586 (2014) 600.

[30] R. Roy Chowdhury, S. Dhara, I. Das, B. Bandyopadhyay, R. Rawat. Journal of Magnetism and Magnetic Materials 451 (2018) 625–628.

[31] K. Momma and F. Izumi, "VESTA 3 for three-dimensional visualization of crystal, volumetric and morphology data," J. Appl. Crystallogr., 44, (2011) 1272-1276.

[32] web site: http://www.materialsproject.org software by computing properties of all known materials.